\documentclass[twocolumn,showpacs,preprintnumbers,amsmath,amssymb]{revtex4}

\usepackage{graphicx}
\usepackage{dcolumn}
\usepackage{bm}

\begin{document}

\preprint{}

\title{Epitaxial growth and transport properties of Nb-doped SrTiO$_{3}$ thin films}

\author{K. S. Takahashi}
 \email{Kei.Takahashi@physics.unige.ch}
\author{D. Jaccard}
\author{J.-M. Triscone}%

\affiliation{DPMC, University of Geneva, 24 Quai Ernest Ansermet, 1211 Geneva
4, Switzerland
}%

\author{K. Shibuya}
\author{T. Ohnishi}
\author{M. Lippmaa}

\affiliation{Institute for solid state physics, University of Tokyo, 515
Kashiwanoha, Chiba 2778581, Japan \\
}%

\date{\today}

\begin{abstract}
Nb-doped SrTiO$_{3}$ epitaxial thin films have been prepared on (001)
SrTiO$_{3}$ substrates using pulsed laser deposition. A high substrate
temperature ($>$1000$^{\circ}\mbox{C}$) was found to be necessary to achieve
2-dimensional growth. Atomic force microscopy reveals atomically flat surfaces
with 3.9 \AA $ $ steps. The films show a metallic behavior, residual
resistivity ratios between 10 and 100, and low residual resistivity of the
order of 10$^{-4}$$\Omega$cm. At 0.3 K, a sharp superconducting transition,
reaching zero resistance, is observed.
\end{abstract}

\maketitle
\section{INTRODUCTION}
\label{sect:intro}  
Electron doped SrTiO$_{3}$ (STO), {\it e.g.} SrTiO$_{3-x}$ and Nb-doped
SrTiO$_{3}$ (Nb-STO), displays superconductivity for carrier densities $n$
between about $1 \times 10^{19}$ and $10^{21}$ cm$^{-3}$ $ $ \cite{J. F.
Schooley,E. R. Pfeiffer,C. S. Koonce}. The critical temperature {\it T}$_{\rm
c}$ strongly depends on $n$ and reaches a maximum of 0.3$-$0.4 K for a carrier
density of about $1 \times 10^{20}$ cm$^{-3}$. In spite of its low {\it
T}$_{\rm c}$, doped STO, being  superconducting with a very low carrier
density, seems to be an ideal model system to exploit the potential
demonstrated in recent conventional field effect experiments and ferroelectric
field effect experiments \cite{C. H. Ahn-1}. Another interesting aspect of STO
is its simple cubic perovskite structure that facilitates the realization of
high quality thin films with flat surfaces. So far, some reports have stressed
that one of the difficulties in synthesizing Nb-doped STO films is to
successfully activate the Nb dopants \cite{K. L. Myers}. Additionally, it is
often found that the film resistivities at low temperatures are higher than
that of single crystals
 \cite{O. N. Tufte,A. Leitner}. Although difficult to obtain, superconductivity in films was reported
for Nb and/or La doped STO in relatively thick films \cite{A. Leitner-2,D.
Olaya}. We previously reported on ferroelectric field effect experiments using
a Nb-STO channel. However, although a clear field induced {\it T}$_{\rm c}$
shift was observed in our field effect devices, zero resistance was not
achieved, even far below the transition temperature, probably as a consequence
of imperfections in the Nb-STO layer or of degradation related to the process
used to fabricate the heterostructure \cite{K. S. Takahashi}. In this paper, we
report on the high temperature and low oxygen gas pressure growth of high
quality epitaxial Nb-doped STO film which allowed us to get full
superconducting transitions in relatively thin films.
\section{EXPERIMENTS AND DISCUSSION}

Epitaxial Nb-STO films were fabricated on (001) SrTiO$_3$ (STO) single crystal
substrates. Pulsed laser deposition (PLD) employing KrF excimer laser pulses
(0.5 J/cm$^{2}$, 1 Hz) focused on a single crystal 1wt\% (2atom\%) Nb-STO
target and very high substrate temperatures were used. In order to reach the
required high substrate temperature of up to 1300$^{\circ}\mbox{C}$, an
infrared (807 nm) diode laser was used, illuminating the substrate from outside
of the UHV chamber through a view port \cite{S. Ohashi}. During the deposition,
reflection high-energy electron diffraction (RHEED) patterns and specular spot
intensities were monitored {\it in situ}. The substrate temperature was kept at
1200$-$1300$^{\circ}\mbox{C}$ under an oxygen pressure of 2$\times 10^{6}$
Torr. After the deposition, the films were cooled down to 500$^{\circ}\mbox{C}$
and kept in 760 Torr of oxygen for 1 hour to re-oxidize both the substrate and
the film. Figure 1 (a) shows an atomic force microscope (AFM) topographic image
of a Nb-STO film, 500 \AA$ $ thick. The scan area is 3$\times$3$\mu$m$^{2}$. No
particles were observed and the straight lines correspond to the 3.9 \AA$ $
height perovskite unit cell steps. Figure 1 (b) shows the specular intensity of
a RHEED reflection as a function of time during the deposition. As can be seen
in the Figure, long period ($\sim$20 s) oscillations and quick recovery, just
after each laser pulse, were observed. The thickness measurements after the
deposition revealed that the long period corresponds to one perovskite unit
cell growth. This result indicates that the growth mode is ``between'' the
layer-by-layer and the step flow growth modes \cite{M. Lippmaa}. The
2-dimensional growth mode is consistent with the flat surfaces measured in AFM
topography images as shown in Fig. 1 (a). Though x-ray diffraction measurements
were performed, no film peak could be detected, the strong substrate peaks very
likely masking the film reflections, suggesting that the lattice constant of
the films is essentially identical to that of the substrate (3.905 \AA).
\begin{figure}
\includegraphics[scale=.5]{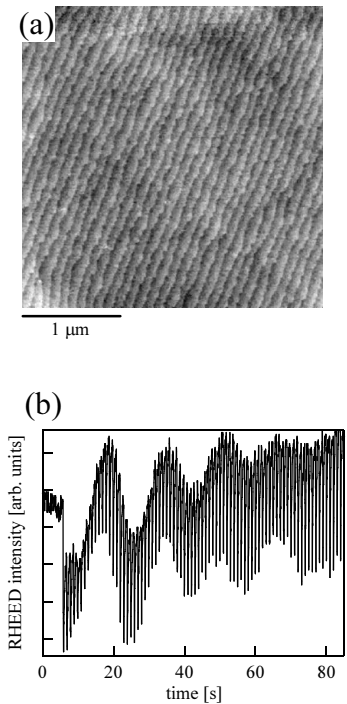}
\caption{\label{fig:1} (a) AFM topographic image of a 500\AA $ $
Nb-doped SrTiO$_{3}$ film. (b) Specular RHEED intensity change
during the deposition.}
\end{figure}
\begin{figure}
\includegraphics[scale=.6]{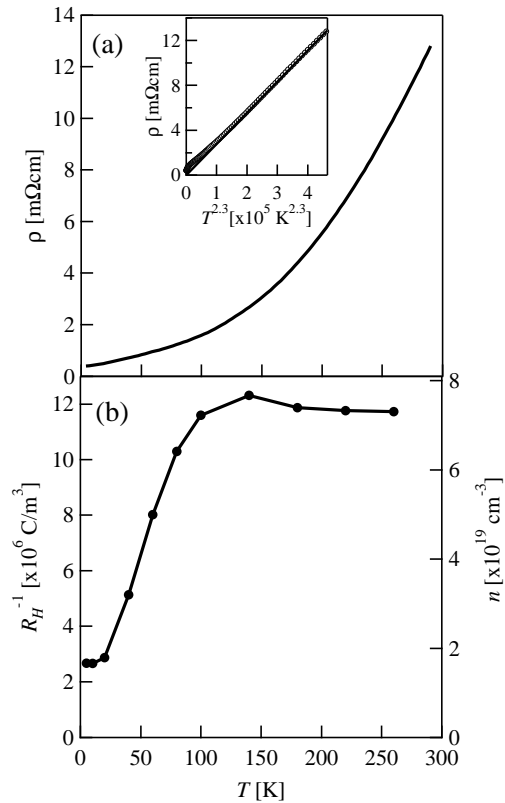}
\caption{\label{fig:2} (a) Temperature dependence of resistivity from 300 K to
4.2 K. The inset shows the $\rho$ vs $T^{2.3}$ plot.  (b) Temperature
dependence of the inverse Hall constant (${R _H} ^{-1}$). Right axis scale is
the carrier density calculated using the simple one band formula (${R _H} ^{-1}
= ne$).}
\end{figure}
To perform detailed transport measurements, photolithography and wet etching
were used to define a conducting path allowing resistivity and Hall effect to
be measured. The current path width was 60 $\mu$m and the distance between the
voltage electrodes 100 $\mu$m. The film thickness of the sample investigated
here is about 260 \AA.\footnote{The measured film was covered by a PZT layer
used for ferroelectric field effect experiments.} Figure 2 (a) shows the
temperature dependence of the resistivity. The room temperature resistivity is
about 13 m$\Omega$cm. Lowering the temperature, one observes a large decrease
of the resistivity, similar to what has been observed in doped STO single
crystals \cite{O. N. Tufte}. The residual resistivity is $\sim$500
$\mu\Omega$cm and the residual resistivity ratio about 26. This low temperature
resistivity value is a record low value for thin films in this doping range.
The lowest low temperature resistivity of Nb-doped single crystal reported is
however one order of magnitude smaller, reaching about 60 $\mu\Omega$cm
\cite{O. N. Tufte}. Though analysis of the temperature dependence of
resistivity was performed, no simple transport mechanism could be identified.
In single crystals a power low $\rho \propto T^{2.7}$, independent of the
doping level, was reported between 150 and 300 K \cite{O. N. Tufte}. Here, as
shown in the inset of Figure 2(a), the resistivity versus temperature is well
described in the same temperature range by $\rho \propto T^{2.3}$, however, the
2.3 exponent cannot be simply related to any scattering mechanism and further
investigations will be necessary to understand transport and the remarkable
metallicity observed in this system. Also, it is quite probable, in view of the
different low temperature resistivity values measured from film to film, that
some structural differences or defects, {\it e.g.} epitaxial lattice strain,
may affect both the resistivity value at low temperature and the temperature
dependence of the resistivity. Figure 2 (b) shows the temperature dependence of
the inverse Hall coefficient (${R _H} ^{-1}$). As can be seen in the Figure,
above 100 K the inverse Hall constant is essentially temperature independent
(as one would expect for a simple metal). However, below 100 K, a dramatic
decrease of the Hall constant is observed with a low temperature value being a
factor of 4 smaller than the high temperature value. This behavior is very
different from the essentially temperature independent Hall constant observed
in single crystals \cite{O. N. Tufte}. The origin of the large decrease of ${R
_H} ^{-1}$ at low temperatures is still to be understood in detail. It could be
related to the existence of two types of carriers in doped STO as documented in
several papers \cite{L. F. Mattheiss_1,G. Binning}. As suggested by recent
ferroelectric field effect experiments, extracting the carrier density from the
${R _H} ^{-1}$ value using the simple relation ${R _H} ^{-1}$=$ne$ is only
possible above 100 K \cite{K. S. Takahashi-2}. Using the high temperature
value, ${R _H} ^{-1}$= $12 \times 10^{6}$ C/m$^3$ , one obtains a carrier
density $n$ of $7.5 \times 10^{19}$ cm$^{-3}$. The ratio between the Nb doping
and the active Nb donors would then only be 22\%, which is less than previously
reported ($\sim $50 \%) for thin films \cite{A. Leitner} and single crystals
\cite{H. P. R. Frederikse,C. Lee}.
\begin{figure}
\includegraphics[scale=.6]{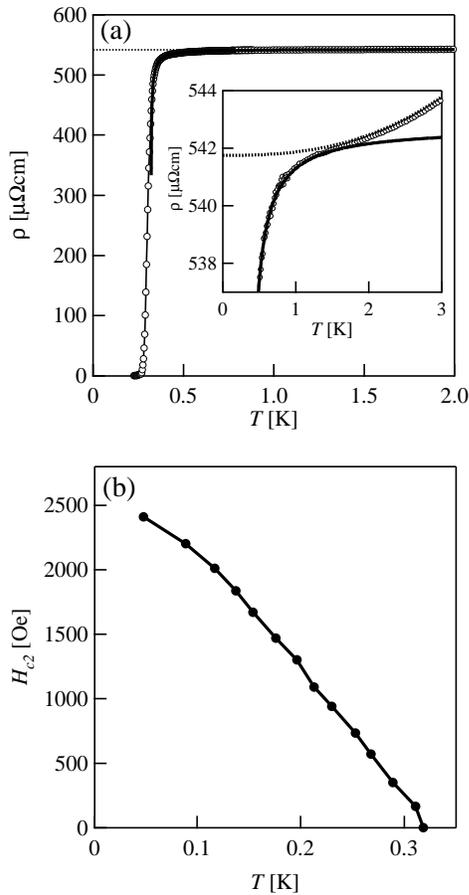}
\caption{\label{fig:3} (a) Temperature dependence of the resistivity below 2 K.
The line is a fit to the curve using a 2D superconducting fluctuation model.
The dotted line is a fit of the normal state from 4.2 K to 1.5 K. The
resistivity curve is well fitted by $\rho _0 + C T ^{2.6}$, although the
mechanism behind this temperature dependence is unclear. The inset shows a
close up of the data from 0.5 K to 3 K. (b) Temperature dependence of the upper
critical field. {\it T}$_{\rm c}$ is defined as the temperature at which the
resistivity decreases to 90\% of the resistivity at 400 mK.}
\end{figure}

Below 4.2 K, a dilution refrigerator was used to perform transport measurements
down to 50 mK. As shown in Fig. 3 (a), the film shows a superconducting
transition around 0.3 K and the resistivity drops to zero at about 0.25 K. We
observe that the resistivity starts to deviate from its normal state behavior
below about 1.4 K as shown in the inset. As shown by the solid line, a
2-dimensional (2D) superconducting fluctuation model (${\it \Delta}\sigma
\propto \frac {T}{T-T_{c}}$)\cite{2D model} allows a good fit to the curve from
1.4 K to just above {\it T}$_{\rm c}$. Superconductivity is suppressed by
applying a magnetic field along the [001] direction (out of the film plane).
Figure 3 (b) shows the upper critical field ($H_{c2}$) as a function of {\it T}
defined as the temperature at which the resistivity reaches 90\% of the normal
state value (taken at 400mK). The temperature dependence can be roughly fitted
to the WHH model that describes conventional dirty limit type-II
superconductors \cite{N. R. Werthamer}. The coherence length at 0 K can be
obtained from the extrapolated zero temperature value of $H_{c2}$ and is about
360 \AA$ $ (using {\it H}$_{c2}$(0) = $\Phi _{0}/2\pi \xi (0)^2$). The fact
that the coherence length is larger than the film thickness 260 \AA$ $
indicates that the system is a 2D superconductor. This is consistent with the
fact that the paraconductivity observed above {\it T}$_{\rm c}$ can be
explained by 2D superconducting fluctuations. Finally, Fig. 4 shows the {\it
T}$_{\rm c}$ vs carrier density $n$ for oxygen reduced and Nb doped single
crystals adapted from references \cite{E. R. Pfeiffer,C. S. Koonce}. Using the
measured {\it T}$_{\rm c}$ value and the carrier density extracted using the
high temperature Hall effect measurements, one can position our thin film on
this graph. As can be seen, the film falls between the two bulk {\it T}$_{\rm
c}$ versus $n$ curves. The reason for the {\it T}$_{\rm c}$ difference between
reduced and Nb-doped single crystals is as yet unknown. It should also be noted
that if the system is a 2D superconductor, one expects a Kosterlitz-Thouless
(K-T) transition with a K-T transition temperature proportional to the
superfluid density (at {\it T}$_{\rm KT}$) \cite{J. M. Kosterlitz}. This should
manifest itself in the tail of the transition and more experiments will be
necessary to precisely determine the exact role played by 2D fluctuations.
\begin{figure}
\includegraphics[scale=.4]{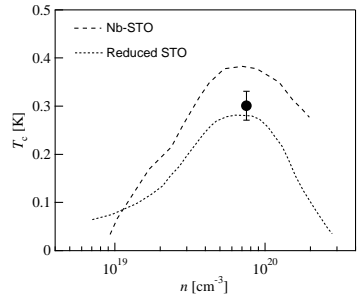}
\caption{\label{fig:4} Dependence of the critical temperature on carrier
density. The behaviors of Nb-doped and reduced STO single crystals are shown as
dotted lines. The point is our Nb-STO thin film.}
\end{figure}
\section{CONCLUSION}
In conclusion, we have fabricated (001) Nb-doped SrTiO$_{3}$ epitaxial thin
films by pulsed laser deposition. Laser heating provided high substrate
temperatures which allowed a 2-dimensional growth to be obtained, as confirmed
by RHEED and AFM measurements. Transport measurements reveal a metallic
behavior with a low residual resistivity and a sharp superconducting
transition. Because of the large coherence length, 2D superconducting
fluctuations are expected and observed close to the superconducting transition.
Such high quality superconducting films might be useful in bringing the
realization of superconducting field effect devices a step closer.

{\bf Acknowledgements}

The authors would like to thank Matthew Dawber for a careful reading of the
manuscript. This work was supported by the Swiss National Science Foundation
through the National Center of Competence in Research, ``Materials with Novel
Electronic Properties, MaNEP'' and division II, New Energy and Industrial
Technology Development Organization (NEDO) of Japan, and ESF (Thiox).


\end{document}